\documentclass[10pt,conference]{IEEEtran}
\IEEEoverridecommandlockouts
\usepackage{cite}
\usepackage{amsmath,amssymb,amsfonts}
\usepackage{graphicx}
\usepackage{textcomp}
\usepackage{xcolor}
\usepackage{siunitx}
\usepackage{tabularx,booktabs}
\usepackage{algorithm}
\usepackage{algpseudocode}
\usepackage{balance}

\begin{document}

\title{Visual Event Detection over AI-Edge LEO Satellites with AoI Awareness}

\author{
\IEEEauthorblockN{Chathuranga M. Wijerathna Basnayaka\IEEEauthorrefmark{1},
Haeyoung Lee\IEEEauthorrefmark{1},
Pandelis Kourtessis\IEEEauthorrefmark{1}, 
John M. Senior\IEEEauthorrefmark{1},\\
Vishalya P. Sooriarachchi\IEEEauthorrefmark{2}\IEEEauthorrefmark{3},   Dushantha Nalin K. Jayakody\IEEEauthorrefmark{2}\IEEEauthorrefmark{3}, Marko Beko\IEEEauthorrefmark{2}\IEEEauthorrefmark{3}\IEEEauthorrefmark{5}
and  
Seokjoo Shin\IEEEauthorrefmark{4}} 
\IEEEauthorblockA{\IEEEauthorrefmark{1}School of Physics, Engineering and Computer Science, University of Hertfordshire, United Kingdom}
\IEEEauthorblockA{\IEEEauthorrefmark{2}COPELABS, Lusófona University, Lisbon 1700-097, Portugal}
\IEEEauthorblockA{ \IEEEauthorrefmark{3}UNINOVA-CTS and LASI, Caparica, 2829-516, Portugal}
\IEEEauthorblockA{ \IEEEauthorrefmark{5} Instituto Superior Técnico, Universidade de Lisboa, 1049-001 Lisbon, Portugal}
\IEEEauthorblockA{ \IEEEauthorrefmark{4}Department of Computer Engineering, Chosun University, Gwangju, South Korea}

\IEEEauthorblockA{Email: \{c.wijerathna-basnayaka-mudiyanselage, h.lee, p.kourtessis, j.m.senior\}@herts.ac.uk, \\ a22211643@alunos.ulht.pt,  dushantha.jayakody@ulusofona.pt, beko.marko@gmail.com, sjshin@chosun.ac.kr}
}

\maketitle

\begin{abstract}
Non terrestrial networks (NTNs), particularly low Earth orbit (LEO) satellite systems, play a vital role in supporting future mission critical applications such as disaster relief. Recent advances in artificial intelligence (AI)-native communications enable LEO satellites to act as intelligent edge nodes capable of on board learning and task oriented inference. However, the limited link budget, coupled with severe path loss and fading, significantly constrains reliable downlink transmission. This paper proposes a deep joint source-channel coding (DJSCC)-based downlink scheme for AI-native LEO networks, optimized for goal-oriented visual inference. In the DJSCC approach, only semantically meaningful features are extracted and transmitted, whereas conventional separate source-channel coding (SSCC) transmits the original image data. To evaluate information freshness and visual event detection performance, this work introduces the age of misclassified information (AoMI) metric and a threshold based AoI analysis that measures the proportion of users meeting application specific timeliness requirements. Simulation results show that the proposed DJSCC scheme provides higher inference accuracy, lower average AoMI, and greater threshold compliance than the conventional SSCC baseline, enabling semantic communication in AI native LEO satellite networks for 6G and beyond.
\end{abstract}


\section{Introduction}
\label{sec:introduction}

The rapid evolution of communication technologies has positioned non-terrestrial networks (NTNs), particularly low Earth orbit (LEO) satellite-assisted systems, as key enablers of mission-critical applications \cite{11126946,10410220}. These includes disaster relief, autonomous vehicles, remote healthcare, industrial automation, all of which demand high reliability, low latency, and fresh, up-to-date information \cite{10412105}.
With the emergence of AI-native communication paradigms, LEO satellites can operate as intelligent edge nodes that perform on board learning and task oriented inference, improving overall network adaptability and efficiency \cite{10694785,9970355}. 
However, LEO satellite communication systems face significant challenges, as their limited link budget, further affected by path loss, rain attenuation, and fading, restricts reliable downlink performance.
Conventional separate source channel coding (SSCC) systems, which process compression and error protection separately, suffer performance degradation under adverse channel conditions. To address these limitations, recent studies have explored deep joint source–channel coding (DJSCC)  \cite{8723589}, a neural network–based framework that jointly optimizes source compression and channel protection in an end-to-end manner. DJSCC has demonstrated strong robustness and graceful performance degradation in noisy environments compared to conventional SSCC systems \cite{yang2022ofdm}. 

As wireless networks move toward semantic and goal oriented communication, the focus shifts from bit level accuracy to task performance, where the receiver must correctly perform tasks such as visual event detection or feature recognition without reconstructing the original data \cite{ma2023theory}. This paradigm is particularly beneficial for satellite systems, where transmission efficiency and information timeliness are critical.
 This paper proposes a DJSCC-based AI-native LEO satellite system for goal-oriented visual inference, transmitting only semantically meaningful features to ground users. To quantify both timeliness and inference accuracy, the age of misclassified information (AoMI) metric is introduced, extending the conventional age of information (AoI) \cite{9380899,10529933} framework to task-oriented semantic communications. Additionally, a threshold-based AoI performance analysis is developed to evaluate system reliability by quantifying the percentage of users meeting application-specific timeliness requirements.

The proposed DJSCC approach is evaluated through simulations in realistic LEO satellite environments characterized by composite fading and rain attenuation. Results demonstrate that the DJSCC scheme significantly improves inference accuracy, information freshness, and threshold compliance rates compared with SSCC baselines, providing a viable foundation for semantic, AI-native satellite networks in 6G and beyond.

\section{System Model}
\label{sec:system_model}

\begin{figure}[!t]
\centering
\includegraphics[width=\linewidth]{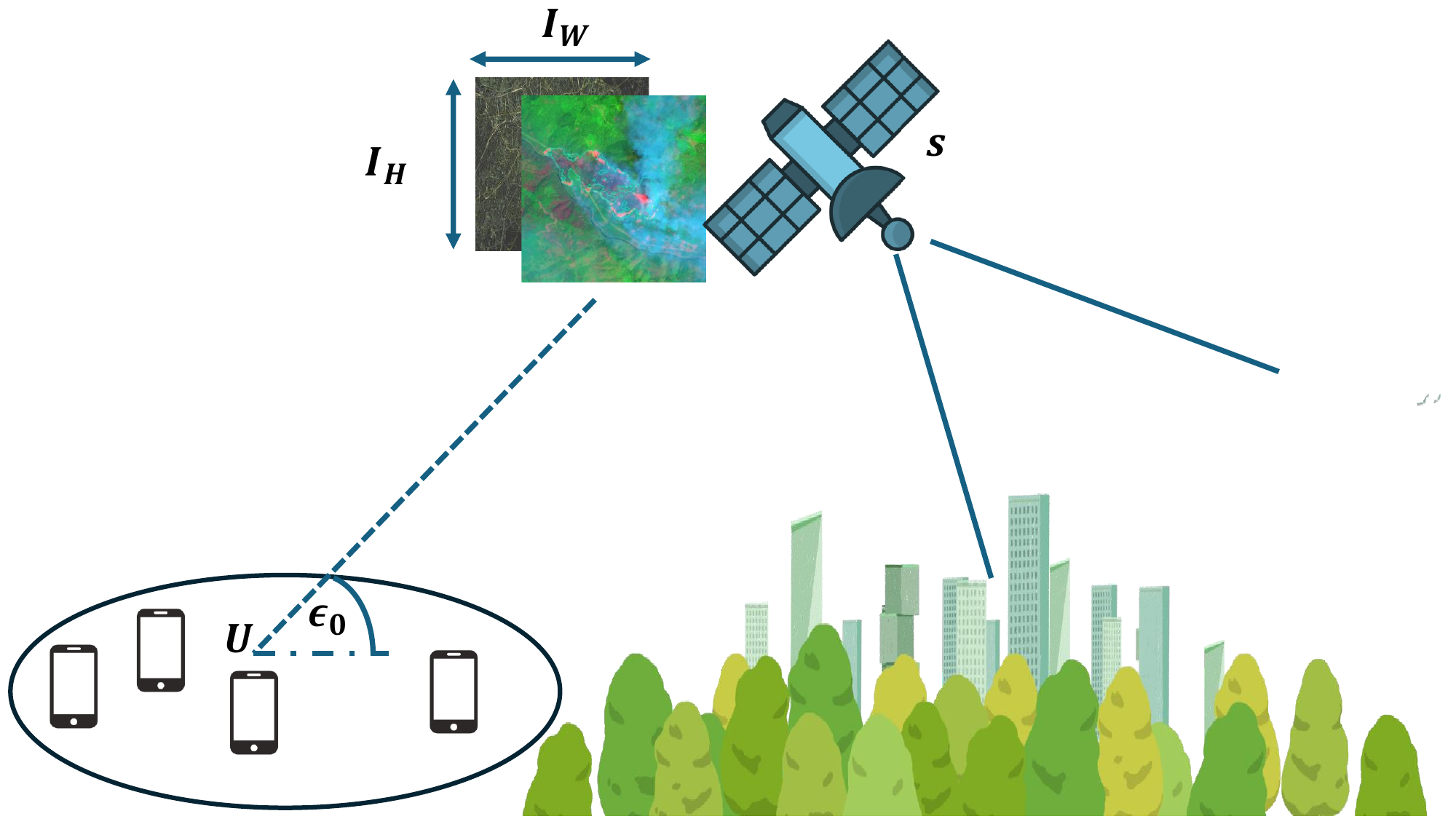}
\caption{System architecture of the proposed LEO satellite communication system. The satellite captures images and transmits them to $U$ ground users (GUs) via DJSCC.}
\label{fig:DJSS}
\end{figure}
\begin{figure*}[!t]
\centering
\includegraphics[width=0.65\linewidth,keepaspectratio]{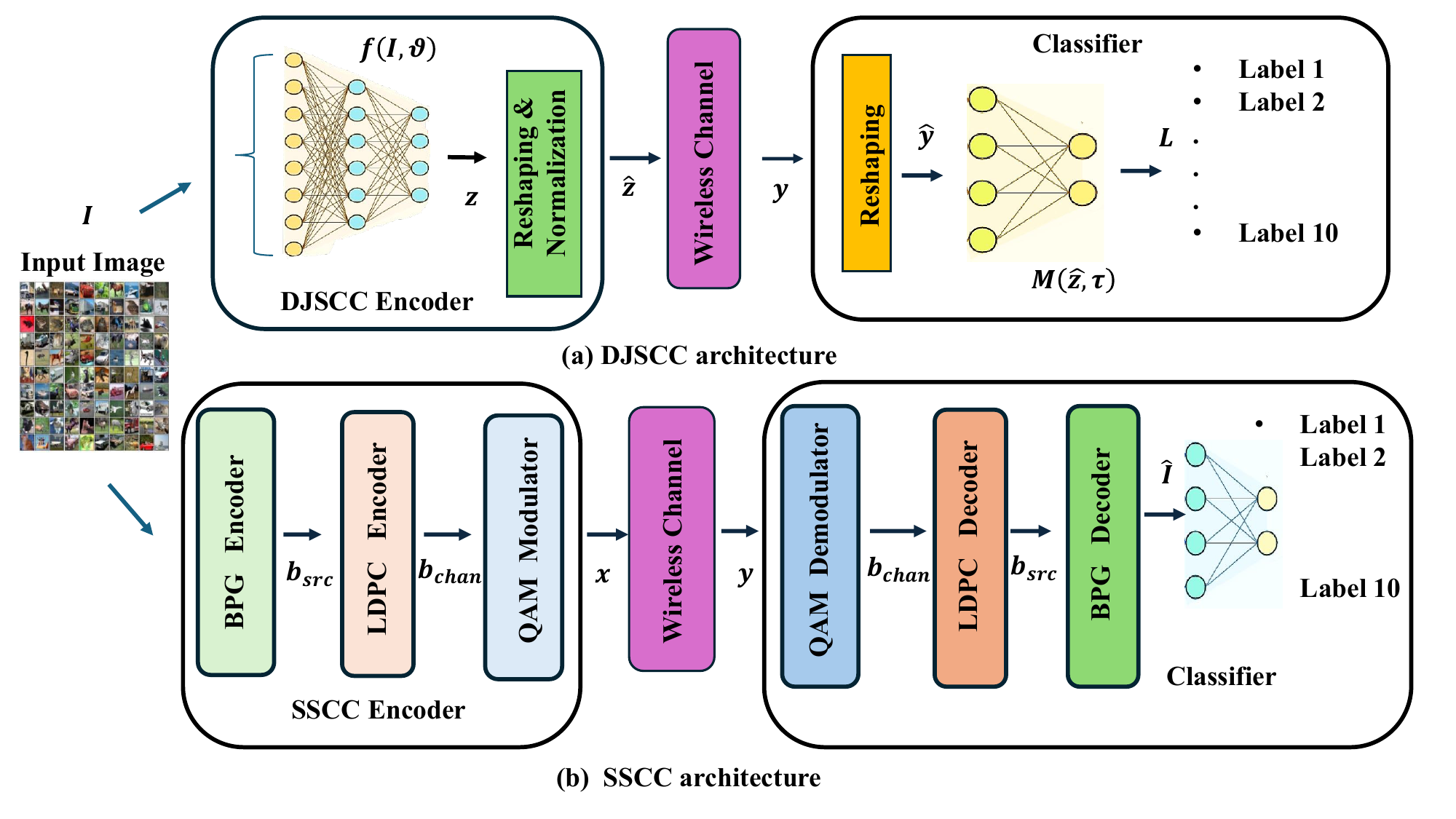}
\caption{(a) DJSCC transceiver; (b) SSCC transceiver. The DJSCC encoder maps images directly to channel symbols; the decoder performs classification without reconstruction. The SSCC chain comprises BPG compression, LDPC coding, QAM modulation, demodulation, LDPC decoding, BPG decompression, and classification.}
\label{fig:receiver_arch}
\end{figure*}
The proposed satellite communication system, illustrated in Fig.~\ref{fig:DJSS}, comprises a single LEO satellite, denoted as $s$, and a set $\mathcal{U}$ of $U$ ground users (GUs). The system is designed for automated visual event detection applications, where the satellite captures images at random intervals during orbital operations, enabling continuous terrestrial monitoring for mission-critical events including natural disasters, environmental changes, and security incidents. This paper proposes a DJSCC scheme that employs an AI-assisted wireless communication framework to facilitate efficient and timely event detection from these randomly acquired images.
Images are captured by the LEO satellite according to a Poisson process with rate $\lambda_I$ images per second. Each captured image $\boldsymbol{I} \in \mathbb{R}^{I_H \times I_W \times I_C}$ undergoes automated processing for event detection. The proposed DJSCC framework enables direct event classification without explicit image reconstruction, as shown in Fig.~\ref{fig:receiver_arch}. The input image has spatial dimensions $I_H \times I_W$ and $I_C$ color channels, comprising $k_P = I_H I_W I_C$ pixels, which defines the source bandwidth. A deep neural network encoder $f(\cdot, \boldsymbol{\vartheta})$ with learnable parameters $\boldsymbol{\vartheta}$ extracts task-specific semantic features $\boldsymbol{z} = f(\boldsymbol{I}, \boldsymbol{\vartheta}) \in \mathbb{R}^{2n_T}$, where $n_T$ denotes the number of transmitted complex symbols. The bandwidth compression ratio $n_T / k_P < 1$ indicates source compression. The encoder produces a latent feature tensor of size $(I_H / 2^\delta) \times (I_W / 2^\delta) \times n_{\text{con}}$, where $\delta \in \mathbb{Z}^+$ is the number of downsampling stages and $n_{\text{con}}$ is the number of feature channels per spatial location. The total number of complex channel symbols is given by
\begin{equation}
n_T = \left( \frac{I_H I_W}{2^{2\delta}} \right) n_{\text{con}},
\end{equation}
yielding the compression ratio
\begin{equation}
\frac{n_T}{k_P} = \frac{n_{\text{con}}}{I_C \cdot 2^{2\delta}}, \quad k_P = I_H I_W I_C.
\end{equation}
The semantic features $\boldsymbol{z} \in \mathbb{R}^{2n_T}$ are  linearly mapped to the transmitted $\boldsymbol{x}_{\text{sem}} = [x_1, \dots, x_{n_T}]^T \in \mathbb{C}^{n_T}$, subject to the average power constraint
$\frac{1}{n_T} \mathbb{E}\left[ \|\boldsymbol{x}_{\text{sem}}\|^2 \right] \leq P_T,$ where $P_T$ is the maximum transmit power. The LEO satellite channel is characterized by composite fading, which integrates large-scale path loss, rain attenuation, and small-scale Rician fading. The received signal for a user $u$ employing DJSCC is modeled as:

\begin{equation}
\boldsymbol{y}_{u} = \sqrt{P_TG_{\mathrm{large}}^{(s,u)}} \cdot \boldsymbol{H}_{s,u} \boldsymbol{x}_{\mathrm{sem}} + \boldsymbol{n}_{u},
\end{equation}
where $G_{\mathrm{large}}^{(s,u)}$ denotes the aggregate large-scale channel gain between satellite $s$ and user $u$, accounting for free-space path loss and rain-induced attenuation. The matrix $\boldsymbol{H}_{s,u} = \mathrm{diag}(h_1, h_2, \ldots, h_{n_T})$ is a diagonal channel matrix whose elements $h_i$ are independent and identically distributed Rician fading coefficients, each representing the small-scale fading experienced by one of the $n_T$ transmitted symbols. The additive white Gaussian noise at the receiver is given by $\boldsymbol{n}_{u} \sim \mathcal{CN}(0,\sigma_{u}^2\boldsymbol{I}_n)$, where $\boldsymbol{I}_n$ is the identity matrix. The small scale Rician fading component models the dominant Line of Sight (LOS) propagation characteristic typical of LEO satellite channels. Each coefficient $h_i$ is expressed as:

\begin{equation}
h_i = \sqrt{\frac{K}{K+1}}h_{\mathrm{LOS}} + \sqrt{\frac{1}{K+1}}h_{\mathrm{NLOS}},
\end{equation}
where $h_{\mathrm{LOS}}$ is the deterministic LOS component, $h_{\mathrm{NLOS}} \sim \mathcal{CN}(0,1)$ represents the scattered non-LOS components, and $K$ is the Rician K-factor, defining the power ratio between the LOS and non-LOS components. The power gain $|h_i|^2$ follows a non-central chi-square distribution. Its probability density function (PDF) is given by:
\begin{equation}
f_{|h_i|^2}(z) = (K + 1)e^{-K} \exp\!\left[-(K + 1)z\right] I_0\!\left(2\sqrt{K(K + 1)z}\right),
\nonumber
\end{equation}
where $z \ge 0$, $I_0(\cdot)$ denotes the zero-order modified Bessel function of the first kind. The total channel power gain is the product of large-scale and small-scale components: $G_{\text{total}}^{(s,u)} = G_{\text{large}}^{(s,u)} |h_i|^2.$ The aggregate large scale channel gain comprises:
\begin{equation}
G_{\text{large}}^{(s,u)} = \frac{G_T G_R}{PL_{\text{total}}},
\label{eq:channel_gain}
\end{equation}
where $G_T$ and $G_R$ represent satellite and terrestrial antenna gains, respectively and $PL_{\text{total}} = \mathrm{PL_{fs}} + \mathrm{PL_{rain}}$ is the composite path loss, which includes free-space loss and rain attenuation. The free-space path loss is
\begin{equation}
\mathrm{PL_{fs}} = \left(\frac{4\pi d f_c}{c}\right)^2,
\label{eq:free_space_pathloss}
\end{equation}
where $d$ is the slant range, $f_c$ is the carrier frequency, and $c$ is the speed of light. The slant range is
\begin{equation}
d = R_E \left( \sqrt{ \left(1 + \frac{o}{R_E}\right)^2 - \cos^2 \epsilon_0 } - \sin \epsilon_0 \right),
\label{eq:slant_range}
\end{equation}
where $R_E$ is the Earth’s radius, $o$ is the orbital altitude, and $\epsilon_0$ is the elevation angle. Rain attenuation is modelled as $\mathrm{PL_{rain}} = \kappa R^\beta L_{\text{rain}}$, where $R$ is the rain rate (mm/h), $\kappa$ and $\beta$ are frequency-dependent coefficients, and $L_{\text{rain}}$ is the effective path length through precipitation: 
\begin{equation}
L_{\text{rain}} = \left[ 0.00741 R^{0.776} + (0.232 - 0.00018) \sin \epsilon_0 \right]^{-1}. 
\end{equation} The signal-to-noise ratio (SNR) for DJSCC-based user $u$ is calculated as
\begin{equation}
\gamma_{u} = \frac{P_T G_{\text{total}}^{(s,u)}}{\sigma_{u}^2}.
\label{eq:snr_semantic}
\end{equation} The receiver processing chain begins with signal reception and preprocessing. The received complex-valued signal $\boldsymbol{y}_{u}$ undergoes component separation, where the real and imaginary components are extracted and reshaped to form the feature vector $\hat{\boldsymbol{z}} \in \mathbb{R}^{2n_T}$. This feature vector serves as input to a deep neural network decoder jointly optimized with the transmitter encoder during the training phase. The decoder network  maps the noisy channel output to semantic features without intermediate bit-level decoding operations. The decoded semantic features $\hat{\boldsymbol{z}}$ are processed by a classification network $M(\cdot, \boldsymbol{\tau})$ with learnable parameters $\boldsymbol{\tau}$, which directly generates the output label $L = M(\hat{\boldsymbol{z}}, \boldsymbol{\tau})$. This approach preserves semantic information and avoids image reconstruction, reducing end-to-end latency.
\par For comparison with conventional approaches, we consider a conventional SSCC scheme. The input image $\boldsymbol{I}$ first undergoes source compression using Better Portable Graphics (BPG), producing a sequence of encoded bits $\boldsymbol{b}_{\text{src}}$. These bits are then channel encoded using a Low-Density Parity-Check (LDPC) code, generating the encoded bit stream $\boldsymbol{b}_{\text{chan}}$. The channel-encoded bits $\boldsymbol{b}_{\text{chan}}$ are modulated using $M$-ary Quadrature Amplitude Modulation (QAM), where $M = 2^m$ represents the constellation size and $m$ is the number of bits per symbol. The modulation process maps each group of $m$ bits to a complex symbol $x \in \mathbb{C}$ from the QAM constellation set $\mathcal{X} = \{X_1, X_2, \ldots, X_M\}$. The transmitted symbol sequence $\boldsymbol{x} = [x_1, x_2, \ldots, x_{N_{\text{sym}}}]^T$ satisfies the average power constraint: $\mathbb{E}\left[\|x\|^2\right] = \frac{1}{N_{\text{sym}}} \sum_{i=1}^{N_{\text{sym}}} \|x_i\|^2 = P_{\text{avg}}$, where $P_{\text{avg}}$ denotes the average symbol power and $N_{\text{sym}}$ is the total number of transmitted symbols. For SSCC scheme, the received signal is given by $\boldsymbol{y}_u = \sqrt{P_{\text{avg}} G_{\text{large}}^{(s,u)}} \cdot \boldsymbol{H}_{s,u} \boldsymbol{x} + \boldsymbol{n}_{u}$, where the channel parameters $G_{\text{large}}^{(s,u)}$, $\boldsymbol{H}_{s,u}$, and the resulting SNR $\gamma_u = {P_{\text{avg}} G_{\text{total}}^{(s,u)}}/{\sigma_u^2}$ are defined identically to the DJSCC case.  The received signal $\boldsymbol{y}_u$ first undergoes symbol detection using maximum likelihood detection, followed by QAM demodulation to recover the encoded bit stream $\boldsymbol{b}_{\text{chan}}$. The demodulated bits are then processed by an LDPC decoder implementing the iterative belief propagation algorithm. The channel decoder corrects transmission errors and outputs the source-encoded bit stream $\boldsymbol{b}_{\text{src}}$, which is subsequently decompressed using the BPG decoder to reconstruct the image $\boldsymbol{\hat{I}}$. The reconstructed image, which may contain distortions from lossy compression and uncorrected symbol errors from the channel decoder, is finally fed into a pre-trained classifier to obtain the classification result. This multi-stage processing pipeline, while providing explicit error correction, introduces significant processing delays compared to the DJSCC approach due to the sequential nature of demodulation, channel decoding, source decoding, and classification operations.

\section{Age of Information Analysis}
\label{sec:aoi_analysis}

This section derives the average AoMI (AAoMI) for the LEO satellite-assisted communication network. The system model assumes images are captured by the satellite at rate $\lambda_I$, following a Poisson process. The probability of correct image classification at receiver $u$ is denoted by $\rho_u$, where $0 \leq \rho_u < 1$ and $u \in \mathcal{U}$. At the transmitter, located on the LEO satellite, the input image undergoes encoding with processing delay $D_{\text{enc}}$. The encoded features are transmitted over the wireless channel with transmission time $n_T T_s$, where $n_T$ is the number of symbols and $T_s$ is the symbol duration. User $u$ performs classification with processing time $D_{\text{cls}}^{(u)}$. Accordingly, the total end-to-end delay experienced by user $u$ is defined as $D_{\text{total}}^{(u)} = D_{\text{enc}} + n_T T_s + D_{\text{cls}}^{(u)}$.

\par For user $u \in \mathcal{U}$, let $\tau_g^{(u)}(t)$ denote the generation timestamp of the most recently correctly classified image at the receiver up to the observation time $t$. The instantaneous AoMI at time $t$ is defined as
\begin{equation}
\alpha_0^{(u)}(t) \triangleq t - \tau_g^{(u)}(t),
\label{eq:instant_aomi}
\end{equation}
which represents the time elapsed since the last successful classification. The time-averaged AoMI over an observation interval $[0, T]$ is given by
\begin{equation}
\bar{\alpha}_T^{(u)} \triangleq \frac{1}{T} \int_0^T \alpha_0^{(u)}(t)  dt.
\label{eq:time_avg_aomi}
\end{equation}
Under ergodicity, as $T \to \infty$, $\bar{\alpha}_T^{(u)}$ converges to the average AoMI (AAoMI):
\begin{equation}
\alpha_{\text{avg}}^{(u)} \triangleq \lim_{T \to \infty} \bar{\alpha}_T^{(u)} = \mathbb{E}\left[\alpha_0^{(u)}(t)\right].
\label{eq:ensemble_avg_aomi}
\end{equation}

The AAoMI is derived by modeling the system using stochastic hybrid systems (SHS) \cite{9380899}. In this framework, the state is defined by a continuous component and a discrete component. The discrete state, $q(t) \in \mathcal{Q} = \{0, 1\}$, represents the operational mode of the system: $q(t)=0$ indicates the system is idle, and $q(t)=1$ indicates it is actively transmitting an image. The continuous state, $\boldsymbol{\alpha}^{(u)}(t) = [\alpha_1^{(u)}(t), \alpha_0^{(u)}(t)]$, tracks the age processes. Here, $\alpha_0^{(u)}(t)$ is the AoMI---the age of the last correctly classified image at the receiver. The state $\alpha_1^{(u)}(t)$ represents the projected age of the image currently under transmission, were it to be correctly classified.

The evolution of the AoMI process is described by a directed graph $(\mathcal{Q}, \mathcal{L})$, where the set of nodes $\mathcal{Q}$ represents the discrete states, and the set of directed edges $\mathcal{L}$ represents transitions between these states triggered by stochastic events. Each transition $l \in \mathcal{L}$ is associated with a transition rate $\lambda^{(l)}$ and a reset map that instantaneously updates the continuous state upon the transition according to $\boldsymbol{\alpha}^{(u)\prime} = \boldsymbol{\alpha}^{(u)} \boldsymbol{A}_{l}$, where $\boldsymbol{A}_{l} \in \{0,1\}^{2 \times 2}$ is a binary matrix.

For the considered system, the transitions are defined as follows. \textbf{Transition \(l_1\) (\(0 \rightarrow 1\))} corresponds to an image generated by the satellite while the system is idle, occurring at rate \(\lambda^{(1)} = \lambda_I\). The reset map \(\boldsymbol{A}_1 = \begin{bmatrix} 0 & 0 \\ 1 & 0 \end{bmatrix}\) sets \(\alpha_1^{(u)\prime} = \alpha_0^{(u)}\) and \(\alpha_0^{(u)\prime} = 0\). \textbf{Transition \(l_2\) (\(1 \rightarrow 0\))} occurs when the ground user correctly classifies a transmitted image, with rate \(\lambda^{(2)} = \rho_u / D_{\text{total}}^{(u)}\). The reset map \(\boldsymbol{A}_2 = \begin{bmatrix} 0 & 0 \\ 0 & 1 \end{bmatrix}\) updates the AoMI by setting \(\alpha_0^{(u)\prime} = \alpha_1^{(u)}\), reflecting the age of the newly correctly-classified image. \textbf{Transition \(l_3\) (\(1 \rightarrow 0\))} represents a misclassification by the ground user, occurring at rate \(\lambda^{(3)} = (1-\rho_u) / D_{\text{total}}^{(u)}\). The reset map \(\boldsymbol{A}_3 = \begin{bmatrix} 0 & 0 \\ 1 & 0 \end{bmatrix}\) resets \(\alpha_1^{(u)}\) while preserving the current AoMI \(\alpha_0^{(u)}\). \textbf{Transition \(l_4\) (\(1 \rightarrow 1\))} is triggered by a new image generated during ongoing transmission, where the current transmission continues and the new image is discarded, occurring at rate \(\lambda^{(4)} = \lambda_I\). The reset map \(\boldsymbol{A}_4 = \begin{bmatrix} 0 & 1 \\ 1 & 0 \end{bmatrix}\) swaps the age values, setting \(\alpha_1^{(u)\prime} = \alpha_0^{(u)}\) and \(\alpha_0^{(u)\prime} = \alpha_1^{(u)}\).  The continuous state $\boldsymbol{\alpha}^{(u)}(t)$ evolves linearly in each discrete state $q$ according to the differential equation: $\dot{\boldsymbol{\alpha}}^{(u)}(t) = \boldsymbol{b}_q$, where $\boldsymbol{b}_q$ is a binary vector. For this system, $\boldsymbol{b}_0 = [1, 0]$ when $q=0$, meaning only the AoMI $\alpha_0^{(u)}(t)$ increases at a unit rate. When $q=1$, $\boldsymbol{b}_1 = [1, 1]$, meaning both $\alpha_0^{(u)}(t)$ and $\alpha_1^{(u)}(t)$ increase at a unit rate.  The SHS method computes the AAoMI, $\alpha_{\text{avg}}^{(u)} = \mathbb{E}[\alpha_0^{(u)}]$, by analyzing the system in steady state. Let $\pi_q(t) = \mathbb{E}[\delta_{q, q(t)}]$ be the probability of the discrete state $q$ at time $t$, and let $\boldsymbol{v}_q^{(u)}(t) = [v_{q0}^{(u)}(t), v_{q1}^{(u)}(t)] = \mathbb{E}[\boldsymbol{\alpha}^{(u)}(t) \delta_{q, q(t)}]$ be the correlation vector between the age process and the discrete state. Assuming the underlying Markov chain is ergodic, the state probability vector $\boldsymbol{\pi}(t) = [\pi_0(t), \pi_1(t)]$ converges to a unique stationary distribution $\bar{\boldsymbol{\pi}} = [\bar{\pi}_0, \bar{\pi}_1]$. The balance equations for the discrete states are 
$\bar{\pi}_0 \lambda_I = \bar{\pi}_1 / D_{\text{total}}^{(u)}, \quad \bar{\pi}_0 + \bar{\pi}_1 = 1$. Solving these yields:
\begin{equation}
\bar{\pi}_0 = \frac{1}{1 + \lambda_I D_{\text{total}}^{(u)}}, \quad \bar{\pi}_1 = \frac{\lambda_I D_{\text{total}}^{(u)}}{1 + \lambda_I D_{\text{total}}^{(u)}}.
\label{eq:stationary_probs}
\end{equation} 

In steady state, the correlation vectors $\boldsymbol{v}_q^{(u)} = \lim_{t \to \infty} \boldsymbol{v}_q^{(u)}(t)$ satisfy the following system of linear equations for all $\bar{q} \in \mathcal{Q}$ \cite{9380899}:
\begin{equation}
\boldsymbol{v}_{\bar{q}}^{(u)} \sum_{l \in L_{\bar{q}}} \lambda^{(l)} = \boldsymbol{b}_{\bar{q}} \bar{\pi}_{\bar{q}} + \sum_{l \in L_{\bar{q}}'} \lambda^{(l)} \boldsymbol{v}_{q_l}^{(u)} \boldsymbol{A}_l,
\label{eq:shs_steady_state}
\end{equation}
where $L_{\bar{q}}$ is the set of transitions leaving state $\bar{q}$, and $L_{\bar{q}}'$ is the set of transitions entering state $\bar{q}$. Applying \eqref{eq:shs_steady_state} to states $q=0$ and $q=1$ gives the following equations. For state $q=0$:
\begin{equation}
\lambda_I \boldsymbol{v}_0^{(u)} = \boldsymbol{b}_0 \bar{\pi}_0 + \frac{\rho_u}{D_{\text{total}}^{(u)}} \boldsymbol{v}_1^{(u)} \boldsymbol{A}_2 + \frac{1-\rho_u}{D_{\text{total}}^{(u)}} \boldsymbol{v}_1^{(u)} \boldsymbol{A}_3.
\label{eq:v0_full}
\end{equation}
Substituting the values for $\boldsymbol{b}_0$, $\boldsymbol{A}_2$, and $\boldsymbol{A}_3$ gives the scalar equations:
\begin{align}
\lambda_I v_{00}^{(u)} &= \bar{\pi}_0 + \frac{1-\rho_u}{D_{\text{total}}^{(u)}} v_{10}^{(u)}, \label{eq:v0_0} \\
\lambda_I v_{01}^{(u)} &= \frac{\rho_u}{D_{\text{total}}^{(u)}} v_{11}^{(u)}. \label{eq:v0_1}
\end{align}
For state $q=1$:
\begin{equation}
\left(\lambda_I + \frac{1}{D_{\text{total}}^{(u)}}\right) \boldsymbol{v}_1^{(u)} = \boldsymbol{b}_1 \bar{\pi}_1 + \lambda_I \boldsymbol{v}_0^{(u)} \boldsymbol{A}_1 + \lambda_I \boldsymbol{v}_1^{(u)} \boldsymbol{A}_4.
\label{eq:v1_full}
\end{equation}
Substituting the values for $\boldsymbol{b}_1$, $\boldsymbol{A}_1$, and $\boldsymbol{A}_4$ gives the scalar equations:
\begin{align}
\left(\lambda_I + \frac{1}{D_{\text{total}}^{(u)}}\right) v_{10}^{(u)} &= \bar{\pi}_1 + \lambda_I v_{01}^{(u)} + \lambda_I v_{11}^{(u)}, \label{eq:v1_0} \\
\left(\lambda_I + \frac{1}{D_{\text{total}}^{(u)}}\right) v_{11}^{(u)} &= \bar{\pi}_1 + \lambda_I v_{10}^{(u)}. \label{eq:v1_1}
\end{align}
The AAoMI can be obtained from the correlation vectors through total expectation as  $\alpha_{\text{avg}}^{(u)} = \mathbb{E}[\alpha_0^{(u)}] = \lim_{t \to \infty} \mathbb{E}[\alpha_0^{(u)}(t)] 
= \sum_{q \in \mathcal{Q}} v_{q0}^{(u)} = v_{00}^{(u)} + v_{10}^{(u)}$. Then, solving  \eqref{eq:v0_0}--\eqref{eq:v1_1} for $v_{00}^{(u)}$ and $v_{10}^{(u)}$, and substituting the stationary probabilities from \eqref{eq:stationary_probs}, yields the closed-form expression for the AAoMI of user $u$:
\begin{equation}
\alpha_{\text{avg}}^{(u)} = \frac{1}{\lambda_I \rho_u} + \frac{D_{\text{total}}^{(u)}}{\rho_u} + \frac{\lambda_I (D_{\text{total}}^{(u)})^2}{1 + \lambda_I D_{\text{total}}^{(u)}}.
\label{eq:aami_individual}
\end{equation}

The network AAoMI is then obtained by averaging the individual AAoMI from \eqref{eq:aami_individual} over all users:
$ \alpha_{\text{avg}}^{\text{net}} = \frac{1}{U} \sum_{u=1}^{U} \alpha_{\text{avg}}^{(u)}.$
The AAoMI enables reliability assessment via threshold-based analysis of information freshness. Let $\eta_{\text{aomi}}$ represent the maximum allowable AoMI for a specific application. For each user $u \in \mathcal{U}$, the threshold compliance condition is defined as: $\alpha_{\text{avg}}^{(u)} \leq \eta_{\text{aomi}}$. The threshold compliance ratio across the network quantifies the percentage of users meeting the timeliness requirement: 
$\Gamma = \frac{1}{U} \sum_{u=1}^{U} \mathcal{I}\left(\alpha_{\text{avg}}^{(u)} \leq \eta_{\text{aomi}}\right), $
where $\mathcal{I}(\cdot)$ is the indicator function that returns unity when the argument is true and zero otherwise.
\section{Numerical Results and Performance Analysis}
\label{sec:numerical_results}
The proposed DJSCC scheme is evaluated through Monte Carlo simulations implemented in Python using TensorFlow and Sionna. The system serves $U=5$ ground users with channel parameters randomly sampled to reflect realistic LEO deployment scenarios. Elevation angles $\epsilon_0$ are uniformly distributed in $[20^\circ, 60^\circ]$, rain rates $R$ in $[0.1, 25]$ mm/h, and Rician $K$-factors in $[5, 15]$ dB. System parameters include: Earth radius $R_E = 6371$ km, satellite antenna gain $G_T \in [28, 32]$ dBi, user antenna gain $G_R \in [23, 27]$ dBi, carrier frequency $f_c = 20$ GHz, orbital altitudes $o \in \{400, 1000\}$ km, noise power $\sigma_u^2 = -99.61$ dBm, and rain attenuation coefficients $\kappa=0.075$, $\beta=1.099$. Transmit power $P_T = P_{\text{avg}}$ ranges from 0.01 W to 100 W.

Performance is assessed using the CIFAR-10 dataset ($32 \times 32 \times 3$ RGB images, $k_P = 3072$ pixels). The SSCC baseline uses BPG compression, rate-$2/3$ LDPC coding with 3072 information bits and 4608 coded bits, and 4-QAM modulation. The BPG quality is adjusted to achieve an end-to-end bandwidth ratio of $1/3$, resulting in 1024 channel uses per image. The DJSCC transceiver adopts the DeepJSCC-$l$ architecture \cite{bourtsoulatze2019deep}, where the encoder
comprises five convolutional layers with generalized divisive normalization (GDN) activations. With $\delta = 2$ downsampling stages, the encoder produces a latent tensor of size $8 \times 8 \times n_{\text{con}}$. Setting $n_{\text{con}} = 16$ yields 1024 complex channel symbols, matching the SSCC bandwidth ratio of $1/3$. The decoder employs transposed convolutions with inverse GDN and a classification head with two fully connected layers. End-to-end training is performed at $\gamma_{\text{train}} = 10$ dB using the Adamax optimizer (batch size 128), cross-entropy loss, and power normalization.

For AoI analysis, encoding delay is $D_{\text{enc}} = 0.01$ s for both schemes, symbol duration $T_s = 125$ ns, classification delays are $D_{\text{cls}}^{\text{(DJSCC)}} = 0.02$ s and $D_{\text{cls}}^{\text{(SSCC)}} = 0.03$ s, and images arrive according to a Poisson process with rate $\lambda_I = 1.0$ image/s.

\begin{figure}[!t]
\centering
\includegraphics[width=0.9\linewidth]{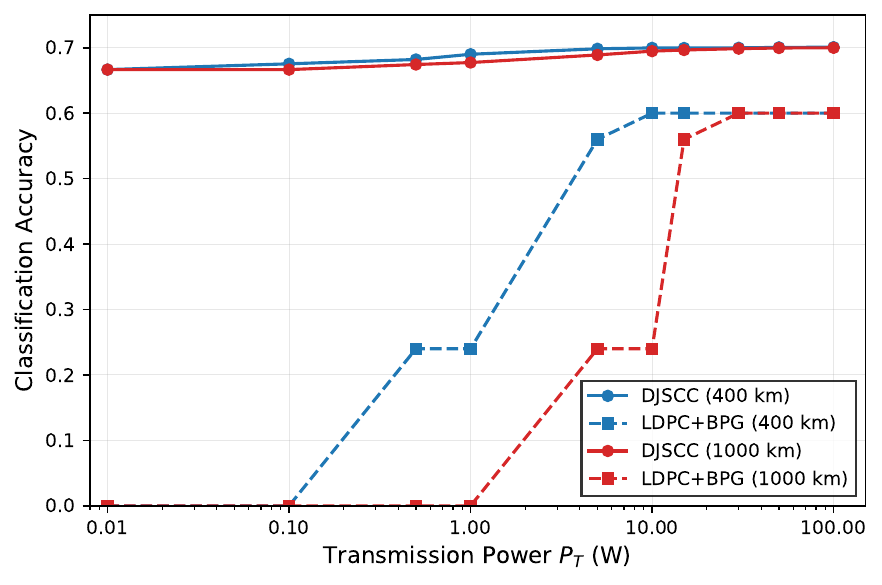}
\caption{Average classification accuracy vs. transmission power for DJSCC and conventional LDPC+BPG at 400 km and 1000 km orbit heights.}
\label{fig:classification_accuracy}
\end{figure}

Fig. \ref{fig:classification_accuracy} illustrates the comparative classification accuracy of DJSCC and conventional LDPC+BPG across varying transmission powers for both 400 km and 1000 km orbit heights. DJSCC achieves higher accuracy than SSCC, especially at low SNR, across both orbit heights. The 400 km orbit achieves slightly better performance owing to reduced path loss compared with the 1000 km orbit. The graceful degradation characteristic of DJSCC is evident, demonstrating the effectiveness of the joint source-channel coding approach for visual event detection in challenging satellite communication environments.

\begin{figure}[!t]
\centering
\includegraphics[width=0.95\linewidth]{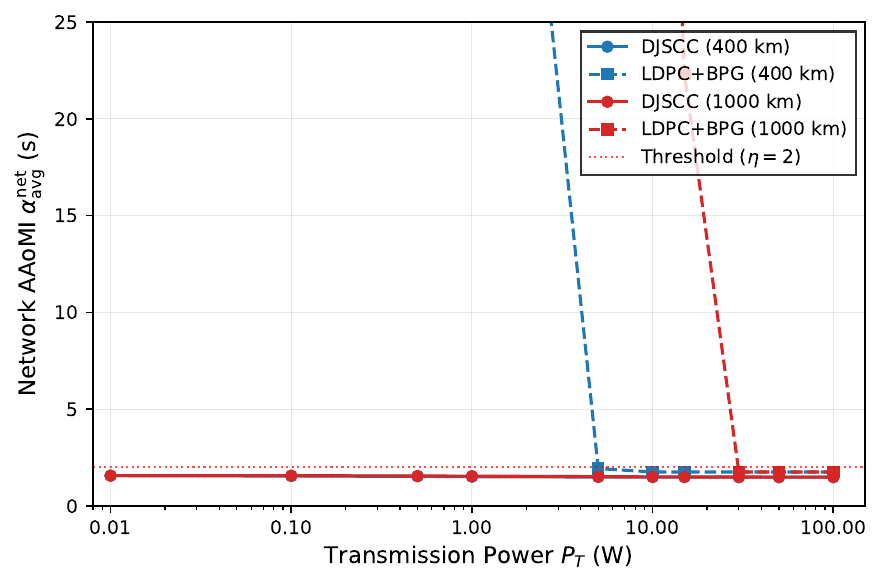}
\caption{Network AAoMI ($\alpha_{\text{avg}}^{\text{net}}$) vs. transmission power characteristics for DJSCC and conventional LDPC+BPG at 400 km and 1000 km orbit heights.}
\label{fig:aami_performance}
\end{figure}

Fig. \ref{fig:aami_performance} presents the network AAoMI ($\alpha_{\text{avg}}^{\text{net}}$) vs. transmission power characteristics for both orbit heights. The results show that DJSCC consistently achieves lower AAoMI values compared to conventional SSCC across all transmission power levels and orbit heights. This performance advantage is attributed to higher classification accuracy and reduced processing latency relative to conventional approaches. The 400 km orbit height demonstrates better AAoMI performance for both methods due to improved channel conditions, but DJSCC maintains its superiority in both scenarios. The DJSCC scheme demonstrates improved freshness performance across multiple SNR levels, with particular advantages over conventional SSCC approaches in challenging channel conditions.

\begin{figure}[!t]
\centering
\includegraphics[width=0.95\linewidth]{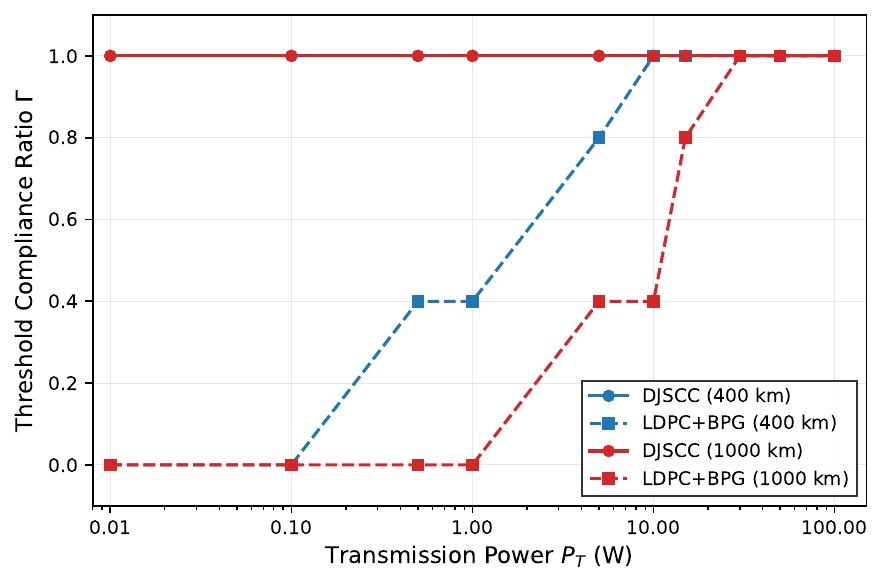}
\caption{Threshold Compliance Ratio $\Gamma$ vs. transmission power characteristics for DJSCC and conventional LDPC+BPG at 400 km and 1000 km orbit heights, with AoMI threshold $\eta_{\text{aomi}} = 2$ seconds.}
\label{fig:threshold_compliance_analysis}
\end{figure}

Fig. \ref{fig:threshold_compliance_analysis} demonstrates the threshold compliance ratio $\Gamma$ across varying transmission powers for both orbit heights, considering an AoMI threshold $\eta_{\text{aomi}} = 2$ seconds. The proposed DJSCC system achieves significantly higher compliance rates compared to conventional SSCC approaches, particularly in low-power operational regimes. For the 400 km orbit height, DJSCC maintains near-perfect compliance ($\Gamma > 0.9$) across most power levels, while conventional SSCC methods struggle to achieve acceptable compliance rates below 10 W transmission power. The 1000 km orbit height shows similar trends with slightly reduced absolute performance due to increased path loss, but DJSCC maintains its relative advantage. Simulation results demonstrate that the DJSCC approach provides advantages over conventional SSCC methods.. For the 400 km orbit height, the DJSCC system achieves up to 287\% relative improvement in classification accuracy at 1~W transmit power compared to the SSCC baseline, highlighting its effectiveness in low-power communication regimes. Furthermore, the proposed DJSCC scheme achieves lower AAoMI across all transmit power levels and reduces AAoMI by more than 85\% at 10 W relative to the conventional SSCC baseline. The threshold compliance analysis reveals that DJSCC achieves compliance rates exceeding 90\% across most operational scenarios, while conventional methods fall below 50\% in challenging conditions.

The performance advantage remains consistent across both orbit heights, with the 400~km orbit achieving approximately 15--20\% superior absolute performance metrics due to reduced path loss. Nevertheless, DJSCC maintains substantial performance gains over conventional methods in both orbital scenarios, demonstrating the robustness of the proposed approach across varying satellite deployment configurations.

While the DJSCC method requires all users to employ AI-enabled processors with substantial computational resources, the significant performance gains in classification accuracy, information freshness, and threshold compliance rates justify this requirement for mission-critical applications. The end-to-end optimization of source compression and channel protection enables robust performance in the challenging LEO satellite environment, making DJSCC a promising approach for future AI-native satellite networks across various orbital configurations.

\section{Conclusion}
\label{sec:conclusion}
This paper introduces a novel deep joint source-channel coding (DJSCC) framework optimised for 6G AI-edge LEO satellite networks supporting visual event detection. The proposed architecture enables direct semantic feature transmission, eliminating explicit image reconstruction while maintaining task-oriented performance. Through comprehensive Stochastic Hybrid Systems analysis, closed-form expressions for the Age of Misclassified Information are derived, providing a unified metric capturing both information timeliness and inference accuracy for semantic communications.
Simulations at 400 km and 1000 km orbit heights show that DJSCC achieves up to 287\% higher classification accuracy at 1 W transmit power, reduces average AoMI by over 85\% at 10 W, and satisfies the timeliness threshold ($\eta_{\text{aomi}} = 2$ s) for over 90\% of users. These performance improvements remain consistent across channel conditions, indicating the effectiveness of DJSCC compared to LDPC+BPG.
The integration of semantic communication with freshness-aware metrics enables reliable low-latency visual inference in AI-native LEO satellite networks, supporting automated environmental and safety monitoring applications such as flood detection, fire surveillance, and urban disaster monitoring.
\balance

\bibliographystyle{IEEEtran}
\bibliography{MyBIB}

@ARTICLE{8723589,
  author={Bourtsoulatze, Eirina and Burth Kurka, David and Gündüz, Deniz},
  journal={IEEE Trans. Cogn. Commun. Netw.}, 
  title={{Deep Joint Source-Channel Coding for Wireless Image Transmission}}, 
  year={2019},
  volume={5},
  number={3},
  pages={567-579},
  doi={10.1109/TCCN.2019.2919300}}

@ARTICLE{9380899,
  author={Yates, Roy D. and Sun, Yin and Brown, D. Richard and Kaul, Sanjit K. and Modiano, Eytan and Ulukus, Sennur},
  journal={IEEE J. Sel. Areas Commun.}, 
  title={{Age of Information: An Introduction and Survey}}, 
  year={2021},
  volume={39},
  number={5},
  pages={1183-1210},
  doi={10.1109/JSAC.2021.3065072}}

@ARTICLE{ma2023theory,
  author={Shao, Yulin and Cao, Qi and Gündüz, Deniz},
  journal={IEEE Trans. Mobile Comput. }, 
  title={{A Theory of Semantic Communication}}, 
  year={2024},
  volume={23},
  number={12},
  pages={12211-12228},
  doi={10.1109/TMC.2024.3406375}}

@ARTICLE{11126946,
  author={Zhao, Xiaohui and Lei, Lei and Wei, Zhiqiang and Fang, Hai and Wang, Wenjie and Chatzinotas, Symeon},
  journal={IEEE Trans. Wireless Commun.}, 
  title={{An Integrated OTFS-NOMA Framework for Multi-Beam LEO Systems: Reliability and Capacity Analysis}}, 
  year={2025},
  volume={},
  number={},
  pages={1-1},
  doi={10.1109/TWC.2025.3594792}}

@ARTICLE{10410220,
  author={Choi, Chang-Sik},
  journal={IEEE Trans. Wireless Commun.}, 
  title={{Modeling and Analysis of Downlink Communications in a Heterogeneous LEO Satellite Network}}, 
  year={2024},
  volume={23},
  number={8},
  pages={8588-8602},
  doi={10.1109/TWC.2024.3351876}}

@ARTICLE{10412105,
  author={Su, Shiying and Jiao, Jian and Yang, Tao and Xu, Liang and Wang, Ye and Zhang, Qinyu},
  journal={IEEE Trans. Wireless Commun.}, 
  title={{Unequal Timeliness Protection Massive Access for Mission Critical Communications in S-IoT}}, 
  year={2024},
  volume={72},
  number={6},
  pages={3211-3226},
  doi={10.1109/TCOMM.2024.3357627},
  ISSN={1558-0857},
  month={June},}

@ARTICLE{10694785,
  author={Yao, Su and Lin, Yiying and Wang, Mu and Xu, Ke and Xu, Mingwei and Xu, Changqiao and Zhang, Hongke},
  journal={IEEE J. Sel. Areas Commun.}, 
  title={{LEOEdge: A Satellite-Ground Cooperation Platform for the AI Inference in Large LEO Constellation}}, 
  year={2025},
  volume={43},
  number={1},
  pages={36-50},
  doi={10.1109/JSAC.2024.3460083}}

@ARTICLE{9970355,
  author={Xiao, Zhenyu and Yang, Junyi and Mao, Tianqi and Xu, Chong and Zhang, Rui and Han, Zhu and Xia, Xiang-Gen},
  journal={IEEE Wireless Commun.}, 
  title={{LEO} Satellite Access Network {(LEO-SAN)} Toward {6G}: Challenges and Approaches}, 
  year={2024},
  volume={31},
  number={2},
  pages={89-96},
  doi={10.1109/MWC.011.2200310}}

@article{yang2022ofdm,
  title={{OFDM-guided deep joint source channel coding for wireless multipath fading channels}},
  author={Yang, Mingyu and Bian, Chenghong and Kim, Hun-Seok},
  journal={IEEE Trans. Cogn. Commun. Netw.},
  volume={8},
  number={2},
  pages={584--599},
  year={2022},
  publisher={IEEE}
}

@article{bourtsoulatze2019deep,
  title={{Bandwidth-agile image transmission with deep joint source-channel coding}},
  author={Kurka, David Burth and G{\"u}nd{\"u}z, Deniz},
  journal={IEEE Trans. Wireless Commun.},
  volume={20},
  number={12},
  pages={8081--8095},
  year={2021},
  publisher={IEEE}
}

@ARTICLE{10529933,
  author={Basnayaka, Chathuranga M. Wijerathna and Jayakody, Dushantha Nalin K. and Perera, Tharindu D. Ponnimbaduge and Beko, Marko},
  journal={IEEE Internet of Things J.}, 
  title={{DataAge: Age of Information in SWIPT-Driven Short Packet IoT Wireless Communications}}, 
  year={2024},
  volume={11},
  number={16},
  pages={26984-26999},
  doi={10.1109/JIOT.2024.3397032}}

\end{document}